# A 20 Gbps Data Transmitting ASIC with PAM4 for Particle Physics Experiments

Li Zhang, Datao Gong, Suen Hou, Guanming Huang, Xing Huang, Chonghan Liu, Tiankuan Liu, Hanhan Sun, Quan Sun, Xiangming Sun, Wei Zhang, and Jingbo Ye

*Abstract*—We present the design principle and test results of a data transmitting ASIC, GBS20, for particle physics experiments. The goal of GBS20 will be an ASIC that employs two serializers each from the 10.24 Gbps lpGBT SerDes, sharing the PLL also from lpGBT. A PAM4 encoder plus a VCSEL driver will be implemented in the same die to use the same clock system, eliminating the need of CDRs in the PAM4 encoder. This way the transmitter module, GBT20, developed using the GBS20 ASIC, will have the exact lpGBT data interface and transmission protocol, with an output up to 20.48 Gbps over one fiber. With PAM4 embedded FPGAs at the receiving end, GBT20 will halve the fibers needed in a system and better use the input bandwidth of the FPGA. A prototype, GBS20v0 is fabricated using a commercial 65 nm CMOS technology. This prototype has two serializers and a PAM4 encoder sharing the lpGBT PLL, but no user data input. An internal PRBS generator provides data to the serializers. GBS20v0 is tested barely up to 20.48 Gbps. With lessons learned from this prototype, we are designing the second prototype, GBS20v1, that will have 16 user data input channels each at 1.28 Gbps. We present the design concept of the GBS20 ASIC and the GBT20 module, the preliminary test results, and lessons learned from GBS20v0 and the design of GBS20v1 which will be not only a test chip but also a user chip with 16 input data channels.

## I. INTRODUCTION

IN particle physics experiments, especially those on the Large Hadron Collider (LHC), high-speed data transmission from the on-detector electronics to the off-detector processing units is challenging. Facing demands such as radiation-tolerance on not only the integrated circuits but also the electrical-optical (E/O) signal converters and the optical fiber, the state-of-the-art in detector data transmission is the lpGBT [1] SerDes ASIC and the VTRx+ [2] optical transceiver working up to 10.24 Gbps per fiber. Both developments (lpGBT and VTRx+) are from the lpGBT and the Versatile Link Plus (VL+) common projects at CERN. The VL+ project also identifies VCSELs (vertical cavity surface emitting laser) at 850 nm for the E/O converter, in either the industrial TOSA (transmitter optical subassembly) LC package, or in die-based array optics, and the matching multi-mode fiber candidates that are measured to be radiation tolerant for applications in the LHC experiments. Given the size of the experiments, the transmission distance and the data rate of 10 Gbps, VL+ uses 150 m as the default length for the selected fibers, which is near the maximum transmission distance of the 10 Gbps NRZ (non-return-to-zero) data rate, with the additional transmission penalties from radiation. At the receiving end of this data link, the off-detector electronics are COTS (components of the shelf) based and follow the advances in industry.

FPGA with embedded serial NRZ and even PAM4 serial data input is the core component in the off-detector data processing units. Limited to 10 Gbps per channel, the fiber counts in a system can be quite high (example, some 40,000 fibers in the ATLAS LAr readout upgrade [3]). The idea of GBS20 grows from the success in LOCx2 [4] in which two serializers share one PLL. We add to that the PAM4 encoder in the same die to eliminate the CDR found in industrial designs and to further simplify the high-speed clocking system. GBS20 output goes directly to a VCSEL, making it possible to build the optical transmitter mezzanine with the serializer ASIC. This module, called GBT20, will remove the requirement of high-speed PCB material for the motherboard which is often large. With all these improvements, the GBS20 ASIC and the GBT20 module are aimed to double the bandwidth in data transmission between on-detector and off-detector electronics, which will either enable new detector development, or reduce cost of existing designs not only in fiber counts, but also the off-detector board counts, crates, and even to simplify the on-detector PCB designs by concentrating the high-speed signals in a pluggable mezzanine.

## II. DESIGN OF THE GBS20 AND GBT20

Shown in Fig. 1 are the diagram of the GBS20 ASIC and a preliminary GBT20 module design in 3D. The work started with the first ASIC prototype GBS20v0. The final goal of this R&D is to use the lpGBT serializer circuit with its full input protocol, and follow the die and chip C4-BGA packaging option. This will save the substrate development cost in this C4-BGA packaging. But the die size of lpGBT (above 20 mm$^2$) is

Li Zhang, Xing Huang, Hanhan Sun, and Wei Zhang are visiting scholars at Southern Methodist University, Dallas, TX 75275 USA. They are PhD candidates at Central China Normal University Wuhan, Hubei 430079, China.

Datao Gong (corresponding author, dgong@mail.smu.edu), Chonghan Liu, Tiankuan Liu, and Jingbo Ye are with Southern Methodist University, Dallas, TX 75275 USA.

Quan Sun is with Fermi National Accelerator Laboratory, Batavia, IL 60510 USA.

Suen Hou is with Academia Sinica, Nangang, Taipei 11529, Taiwan

Guangming Huang and Xiangming Sun is with Central China Normal University Wuhan, Hubei 430079, China.

too large (hence too expensive) for early prototypes. Because of this, GBS20v0 contains only the core circuits. Without data input, we managed to fit all the circuits into a die of 2 mm$^2$. The second prototype GBS20v1 will be 4 mm$^2$, still much smaller than lpGBT, will have user data input at 1.28 Gbps to fit into the available number of pads without going to a C4-BGA pad arrangement. This is a balance between usefulness of this prototype and the overall development cost. GBS20v1 will be submitted in Nov. 2020. The challenge in the design of the mezzanine transmitter GBT20 lies in the electric I/O if we keep the full input scheme of lpGBT with the e-link data rates from 320 Mbps per channel to 1.28 Gbps. By going to a 0.8 mm p-BGA with an interposer [6], this mezzanine can be easily tested, stay as a pluggable, or be soldered on a motherboard.

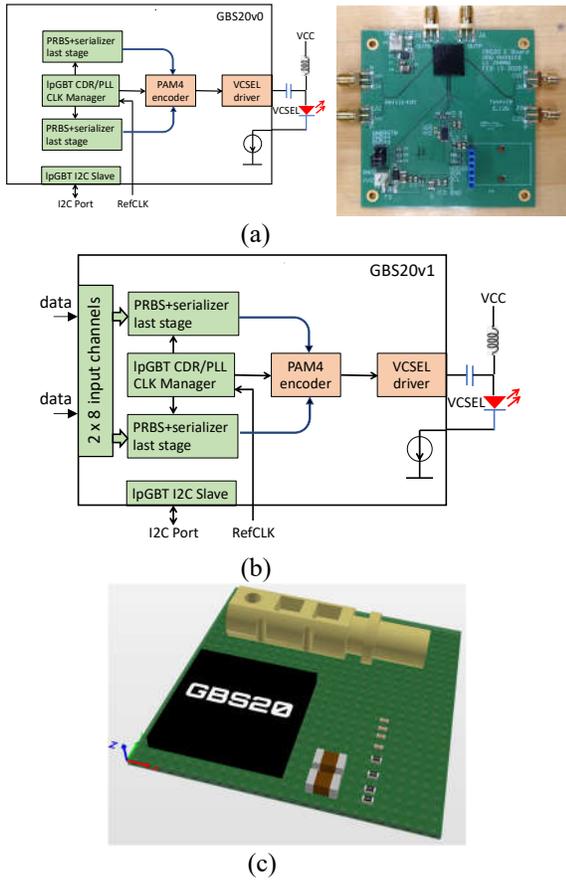

Figure 1. (a) The block diagrams of GBS20v0 and its test PCB board; (b) The block diagram of GBS20v1; (c) a 3D model of GBT20, the transmitter mezzanine card, the bottom of this card is p-BGA.

### III. TEST OF GBS20V0 AND DESIGN IMPROVEMENTS IN GBS20v1

GBS20v0 is fabricated in a 65 nm commercial CMOS technology and tested up to 20 Gbps. The measured PAM4 electric output is shown in Fig. 2. The power supply is chosen to be 1.2 V and this is found to limit the dynamic range of the PAM4 driver. When we set a fixed 2:1 ratio of the MSB and LSB inputs to the encoder, we observe significant nonlinearity in the MSB amplitude, which is partially due to a mistake in the layout. The CTLE technique further cuts into the dynamic range of the MSB amplifier. Because of these design issues, plus a few other mistakes in the layout that contribute to power noise, we decide not to continue the test with a VCSEL. Instead we move on to the next ASIC prototype design work.

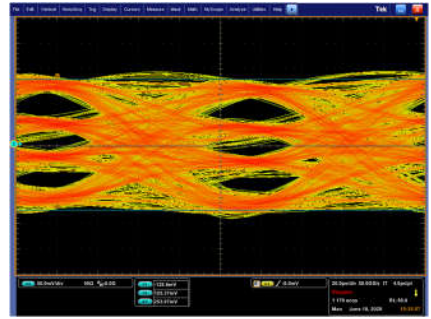

Figure 2. The PAM4 electric output from GBS20v0.

Taken the lessons from GBS20v0, we are implementing the following changes in GBS20v1: the PAM4 stage will operate under 2.5 V to increase the modulation current and leave more headroom for the CTLE bandwidth extension. The gain control to the MSB and LSB amplifiers will be independent from each other. The inductance peaking in the output driver is stronger than before. A programmable source load is added to remove the potential overshot due to the peaking, as shown in Figure 3 (a). These design changes allow for an increase of the maximum output current from 10 mA to 18 mA with better simulated eye diagram, as shown in Figure 3 (b). We plan to submit this prototype in November 2020.

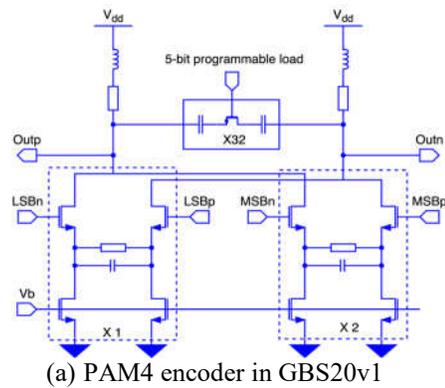

(a) PAM4 encoder in GBS20v1

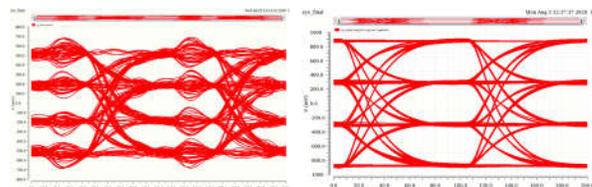

(b) GBS20v0 at 10 mA, GBS20v1 at 18 mA

Figure 3. (a) The PAM4 encoder in GBS20v1, with the programmable source load; (b) Simulated eye from GBS20v0 at 10 mA and GBS20v1 at 18 mA. With properly adjust the load to match the transmission line, the ring is removed.

## IV. Conclusion

We present the new concepts employed in the R&D of the GBS20 ASIC and the GBT20 transmitter mezzanine with a goal to reach 20.48 Gbps per fiber using PAM4 for detector data transmission in particle physics experiments. Test results from the first prototype indicate a speed barely at 20.48 Gbps and revealed several design faults that are to be corrected in the second prototype. Taken the lessons from GBS20v0, we are working on the next prototype GBS20v1, with an aim to submit in November 2020. This R&D using a 65 nm CMOS technology may also serve as guineapig of PAM4 in radiation environment for higher data rates using more advanced CMOS technologies in the future.

## Acknowledgment

We thank Szymon Kulis, Jeffery Prinzie and Paulo Moreira for the permission of using the serializer, the PLL, the I$^2$C and several other design-blocks in lpGBT in this R&D work. We also thank financial support from the SMU University Research Council Grant, the Dedman College Dean's Research Council Grant, and from the Institute of Physics, Academia Sinica.